# A New Approach to Manage QoS in Distributed Multimedia Systems

Bechir Alaya, Claude Duvallet, Bruno Sadeg
LITIS, UFR des Sciences et Techniques
25 rue Philippe Lebon, BP 540
F-76058 Le Havre Cedex
Firstname.Lastname@litislab.fr

*Abstract*—Dealing with network congestion is a criterion used to enhance quality of service (QoS) in distributed multimedia systems. The existing solutions for the problem of network congestion ignore scalability considerations because they maintain a separate classification for each video stream. In this paper, we propose a new method allowing to control QoS provided to clients according to the network congestion, by discarding some frames when needed. The technique proposed, called (m,k)-frame, is scalable with little degradation in application performances. (m,k)-frame method is issued from the notion of (m,k)-firm real-time constraints which means that among k invocations of a task, m invocations must meet their deadline. Our simulation studies show the usefulness of (m,k)-frame method to adapt the QoS to the real conditions in a multimedia application, according to the current system load. Notably, the system must adjust the QoS provided to active clients[1] when their number varies, i.e. dynamic arrival of clients.

## I. INTRODUCTION

Packet switched networks are increasingly being utilized for carrying real-time traffic which often requires quality of service (QoS) in terms of delay, jitter, and loss of packets. A particular type of real-time traffic is a real-time stream, in which a sequence of related packets arrives at regular intervals with certain common timing constraints.

These applications deal with large volumes of data and require real-time processing, i.e., they must be completed before fixed dates, to guarantee an acceptable QoS in the streams presented to users. Systems suitable to the management of these kinds of data with QoS guarantees are real-time database systems (RTDBSs) [1][2].

Besides, many distributed multimedia applications must face to unpredictable loads that cause the system overload. For example, user-demands may arrive in a bursty manner during a short period. Currently, all applications need to provide an acceptable QoS to the users (a good flow of video frames). To this end, we are interested in the adaptation of existing techniques in RTDBSs for multimedia applications in order to obtain more reliable and more efficient transfer of the video packets, without modifying the initial infrastructure.

The main problems are related to the adaptation of available resources (bandwidth, buffer size, video servers, etc.) and to the proposition of new techniques which deal with system instability periods (overload or under-utilization). The proposition must allow to ensure an acceptable QoS while respecting the multiple requirements of the video streams.

An example of multimedia applications is a video-on-demand or a streaming audio. It is important for the application to receive and process the information at an almost constant rate, eg., 30 frames per second for video information. However, due to the network problems, some packets of a video frame can be lost, resulting in little or no noticeable degradation in the QoS at the receiver. More concretely, we consider the MPEG transmission where some frames containing control information (for example synchronization) are inserted into the packet stream in regular way. The video packet can tolerate a certain deadline miss rate only if the deadline misses are uniformly distributed. A large number of consecutive deadline misses cannot be acceptable. Therefore, we should provide adaptive mechanism for controlling deadline miss distribution, to achieve graceful performance degradation [3][4][5].

Many works on QoS management in RTDBSs have been done [6][7]. Almost all these works are based on a feedback control scheduling architecture (FCSA) that controls the system behavior thanks to a feedback loop.

The feedback loop begins to measure the performances of the system in order to detect overload periods. Then, according to the results observed, the values of the parameters are modified to adjust the system load to the real conditions. As these conditions always vary, this process is repeated indefinitely.

Because of the similarities existing between RTDBS and multimedia applications [8], in this paper, we propose to apply the results obtained on the QoS management in RTDBSs to multimedia applications.

The main objective is to allow to design multimedia applications that will be able to provide the QoS guarantees and a certain robustness when user's demands quickly grow up leading to the network congestion. These works are especially applied to video on demand (VoD) applications. The remainder of this paper is structured as follows. In section 2, we present the multimedia system architecture that we use. In section 3, we describe our approach which allows to increase the applications QoS during overload periods (network congestion). In section 4, we propose to integrate (m,k)-frame constraints to provide bandwidth guarantee. Section 5 concludes this paper.

---
[1] Clients that are sending requests.

## II. RELATED WORKS

### A. Management of QoS in real-time database systems

We consider firm RTDBS model, in which late transactions are aborted because they are useless after their deadline, and we consider a main memory database model. This work on QoS guarantees is guided by the following premises:

1) Transactions are executed according to their priority, i.e. a high priority transaction preempts a lower priority transaction, and they are classified into two categories: update transactions and user transactions (see section II-A1).
2) We keep different versions for each data item. These versions are dynamically adjusted by checking the data freshness and by considering Data Error (DE) parameter. Data Error is computed by comparing the data version stored in the database with the corresponding value of the data in the real world. DE must be less than or equal to an upper bound given by MDE parameter[2], related to the data [9]. In the real-time database, validity intervals are used to maintain temporal consistency between the real world values and the sensor data stored in the database [2]. A data version $d_i$ is considered temporally inconsistent (not fresh) or stale if the current time is later than the timestamp of $d_i$ added to the absolute validity duration of $d_i$ (see next paragraph).

*1) Data model and transaction model:* In real-time database, data objects are classified into real-time or not real-time data. A not real-time data is classical data found in conventional databases, whereas a real-time data has a validity duration beyond which it becomes useless. These data change continuously to reflect the real world state (e.g. the current temperature value). Each real-time data has a timestamp representing the last update of the data, i.e. the instant of the last observation of the real world state. Many versions of a real-time data item may be stored in the database and the number of versions may be either fixed or dynamically adjusted. Storing a data version is done according to data freshness and MDE parameters.

Transactions are classified into two classes: update transactions and user transactions. Update transactions are used to update the values of real-time data (sensor data) in order to reflect the real world state. Update transactions are executed periodically and have only to write sensor data. User transactions, representing user requests, arrive aperiodically and may only read real-time data, and/or write not real-time data.

*2) General model:* It is well known that feedback control approach is very effective in management of QoS in RTDBS, under unpredictable workloads [9]. The goal is to control the system performances, defined by a set of controlled variables, in order to satisfy a given QoS specification. The general outline of the feedback control scheduling architecture is given in Figure 1. A RTDBS consists of several components. For the QoS management, a Monitor, a miss ratio controller, an utilization controller and QoD (Quality of Data) manager are added to the system in order to adjust its performances and to control the information flows. An *Admission Controller* (AC) is used to avoid system overload by rejecting some user transactions, if needed. A transactions handler, which provides a platform for managing transactions, consists of a *Concurrency Controller (CC)*, a *Freshness Manager (FM)* and a *Basic Scheduler (BS)*. Transactions are scheduled by a Basic Scheduler in the ready queue, by using, for example, EDF scheduling policy [10]. The FM checks the freshness of a data item before a transaction accesses it. It blocks a user transaction if the target data item is stale. Based on the two-phase locking principle, CC ensures the concurrent transactions serializability. In case of conflict between transactions, when a higher priority transaction uses the data item, transactions with lower priority will be blocked. At each sampling period, the Monitor samples the system performance data from the transaction manager and sends them to the controller. The utilization controller, using the miss ratio, generates a signal based on the sampled miss ratio and utilization data. Feedback control has been proven to be very effective in supporting a required performance specification [9].

### B. Quality of service in distributed multimedia systems

QoS in a multimedia application represents the whole requirements in terms of bandwidth, quality of visualization, delay and rate of video packets loss. Our approach consists in taking into account researches already done on the management of QoS in RTDBSs [11][9] and researches on (m-k)-firm constraints in RTSs [12] and RTDBS s[13], and to adapt them to distributed multimedia systems. The method, we propose, is based on both feedback control architecture for distributed multimedia systems [8] and the notion of (m,k)-firm constraints [14].

This adapted method is called FCA-DMS (Feedback Control Architecture for Distributed Multimedia Systems). We apply a control of the network congestion by discarding or not some multimedia frames of certain types (see next paragraph) according to the network state, notably to the shared bandwidth. This increases the QoS provided to users.

### C. Feedback control architecture

In a previous work, Natalia Dulgheru has proposed an architecture, named QMPEGv2 [8], which deals with distributed multimedia systems (cf. Figure 2).

---

[2] Maximum Data Error.

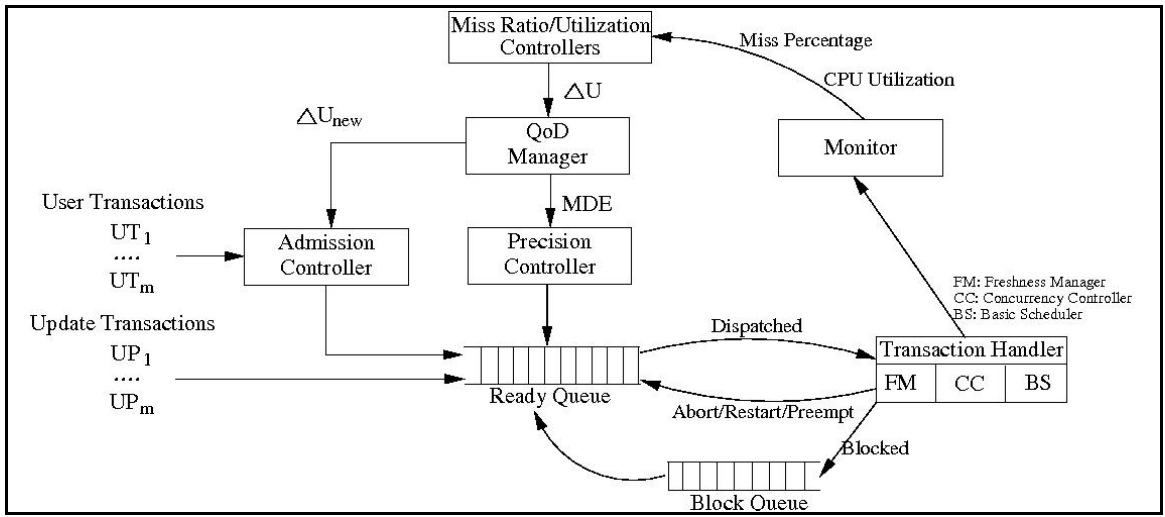

Figure 1. Feedback control scheduling architecture for RTDB.

The architecture proposed contains three main components:

- **A master server:** it accepts requests from clients, chooses the video servers able to serve the demand, supervises the system state and adjusts the video streams in order to maintain the QoS initially fixed.

- **Video servers:** they run under the control of the master server and send the video packets to the clients.

- **Clients:** they send requests to the master server and receive the video frames from the video server. When a state change occurs, they send a feedback report to the master server.

In the following, we describe briefly a typical procedure executed when a video-on-demand is requested, based on FCA-DMS architecture:

1) *A client sends a request to the master server to get a video, with a certain level of QoS.*
2) *The master server broadcasts the request to the video servers available in the system.*
3) *The video servers send back their response to the master server, which chooses one server among them.*
4) *A stream is opened between the chosen video server and the concerned client.*
5) *The master server asks the video servers to adapt their QoS, when necessary.*

The feedback loop consists on adapting the QoS according to the load system conditions (servers and network congestion). The system observes the QoS obtained by the client and, if necessary, asks the concerned video server to improve it.

In order to improve the QoS, the system increases or decreases the number of transmitted frames of certain types (see below). To this purpose, we based our work on the characteristics of the standard MPEG format [15], that defines a mechanism to code frames simultaneously to the video compression.

When a video sequence enters the system, it is compressed and coded according to three types of frames: *Intra frames* (I), *Predicted frames* (P) and *Bidirectional frames* (B). *I* frames are references frames. *P* frames allow to rebuild a frame using an *I* frame. *B* frames use both *I* frames and *P* frames to rebuild a sequence. Therefore, *I* frames are the most critical in the system. To decrease the eventual network congestion, it is necessary to remove some frames from a video sequence in a controlled manner. We propose in the following section a method based on the controlled frames suppression in order to control the QoS provided to users.

### D. Feedback control loop

Using the feedback loop allows to stabilize the system during the instability periods [16]. It is based on observation and auto-adaptation principles.

The observation principle consists of observing the results obtained by the system and checking if the current QoS observed is consistent with the QoS initially required, e.g. in VoD application, the system checks if the video sequences are presented to users without interruptions.

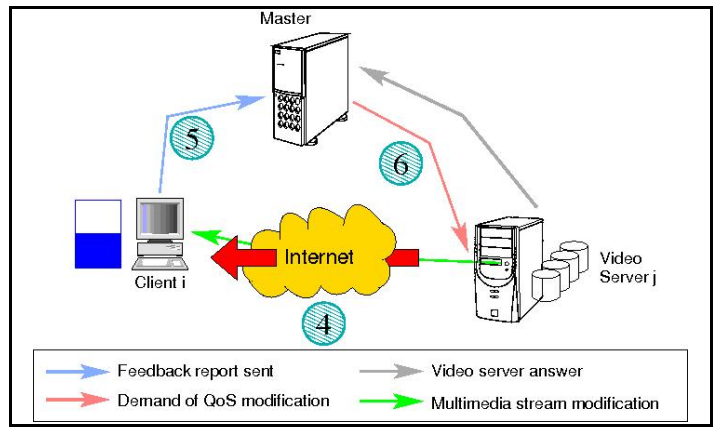

Figure 2. Adapted feedback loop for multimedia applications.

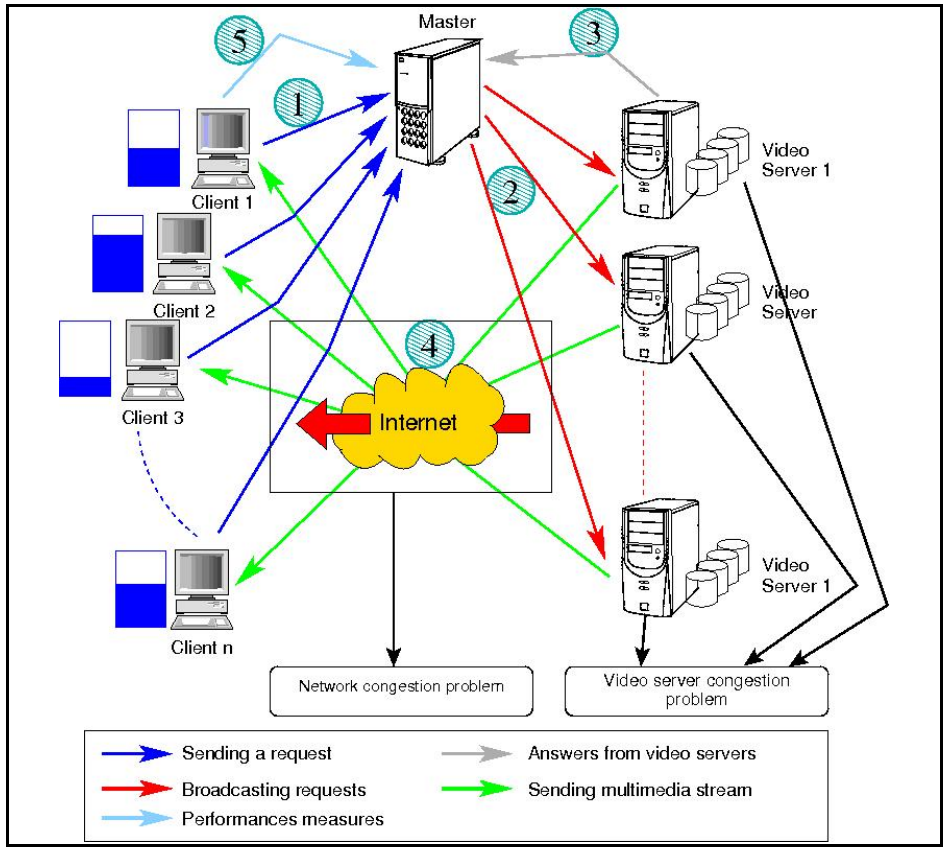

Figure 3. Feedback control architecture for distributed multimedia systems.

The auto-adaptation consists for the system to adapt the results according to the QoS needed by the clients, by adjusting some network and video parameters, e.g. the system increases or decreases the number of accepted frames[3]. This way, the feedback loop ensures the stability of the system.

### III. (M,K)-FRAME METHOD TO CONTROL THE NETWORK CONGESTION

According to certain conditions, the system load varies from overload state to under-utilization state and vice-versa. Indeed, since the number of video servers sending the video packets is unknown, sometimes this causes severe damages on the service level provided to clients. Consequently, the number of transmitted packets is also unknown and can be important. Moreover, when a high number of video packets access to network resources, it is necessary to keep a high priority level for more critical packets (I frames, then B frames, then P frames) [8][17].

We propose an approach based on (m,k)-firm method [14] to maintain a certain QoS in distributed multimedia systems by selecting frames to delete in a controlled manner.

---

[3] Note : I frames (critical) are not removed.

#### A. The (m,k)-firm method

The concept of graceful QoS degradation refers to the degradation of system performance in such a way that the system continues to operate for providing an acceptable reduced level of service. In an overload situation, the QoS degradation is unavoidable since packets will always be delayed or dropped. However, many streams can tolerate some deadline misses if they occur in an accepted manner. It has been shown in [18], that QoS of an audio stream is only sensitive to the consecutiveness of deadline misses. Moreover, Boyce and Gaglianello [19] stated that the effect on QoS of the video streams depends on when and how the loss occurs.

Hamdaoui and Ramanathan [14] formulated, under the concept of the (m,k)-firm model, the ability of a real-time stream to miss some deadlines without degrading drastically the QoS. The method which consists to adapt (m,k)-firm constraints to multimedia applications, that we call (m,k)-frame method, can be stated like the following: a stream is said to have an (m,k)-firm requirement if at least m packets inside any window of k consecutive packets meet their required deadlines. If more than $(k - m)$ deadline misses occur in a specified window of k consecutive packets, then the stream is said to be in a dynamic failure state, i.e. its QoS constraints is not satisfied.

## B. Quality of service adaptation

A video stream is decomposed into several classes according to their tolerance to the loss of frames characteristics, i.e. each class contains the video packets of similar (m,k)-frames constraints. We consider three classes of frames: I, B and P. With this technique, we realize a trade-off between the shared resources and the QoS granularity in the same class of a video stream.

In this work, we focus on the adaptation of the video stream to the network state. We assume that measures of the network capacity are available, in one hand, and that we have an important number of frames to send, on the other hand.

The three classes of frames (I, B and P) are used to adapt the quality of stream sent to the network capacity. We consider the following constraints: $(m_I, k_I)$-frame, $(m_P, k_P)$-frame and $(m_B, k_B)$-frame, i.e, $m_I$ frames of a certain type must be received among the $k_I$ frames sent. Then the network capacity is measured by the formula: $m_I + m_P + m_B$. Remember that I frames are the most critical. The parameters are ordered in the following manner: $m_I > m_P > m_B$. We usually have $m_I = k_I$, i.e., I frames are critical and it is forbidden to remove them.

We assume the situation where the network, whose current capacity is N, is congested. We also assume that QoS$_{max}$ is the quality of the stream to send including M frames. To be consistent with the network capacity, it is necessary to remove (M − N) frames. Therefore, we have to degrade the quality of the MPEG stream. When we apply no method of congestion control, frames will be randomly removed, i.e. they are lost by the network, causing the degradation of the video presentation, notably if some I frames are removed. Here, we apply our (m,k)-frame method, which consists of removing frames in an intelligent manner. We have: (1) M=$k_I + k_P + k_B$, and (2) N=$m_I + m_P + m_B$. The number of frames to remove is then: M − N = $(k_I − m_I) + (k_P − m_P) + (k_B − m_B)$, where $k_I = m_I$ (I frames are the most critical, and are not to remove).

## C. Bandwidth fair sharing

With the previous assumptions, we tackle the problem of sharing network bandwidth between servers in case of congestion phases.

In the previous section, we have seen how to reduce the QoS at the stream level, according to the available capacity of the network. Here, we need to share as fairly as possible the bandwidth between all sources that wish to send a stream.

We begin to compute the total capacity needed by all servers. Then, we compute R, the ratio between the needed capacity and the available network capacity (N).

$$R = \frac{N}{\sum_{i=1}^{m} RC_i} \quad (1)$$

where:
- *m*: number of video server.
- $RC_i$: required capacity of the video server *i*.

Example: let 3 video servers wishing to send flows of 40, 30 and 20 frames per second respectively. The total capacity of the network needed to answer to this demand must be 40+30+20=90 frames per second. If the network only arranges a capacity of 75 frames per second, then it will not be able to send all the frames. We compute the ratio R as follows: (75/90)×100 = 83.33%. Then, we apply this rate to each of the three required capacities 40×83.33%=33, 30×83.33%=25 and 20×83.33%=17. When we sum the three obtained numbers, we find 75 frames per second. This corresponds to the actual capacity of the network.

*Algorithm 1:* (m,k)-frame algorithm

```
begin
    selection of a video stream T;
    selection of a GoP[i] in T /* GoP = Group Of Pictures */
    if (GoP[i]='I') then
        mark the frame as mandatory
    else if (GoP[]='B') then
        mark the frame as hard optional
    else if (GoP[]='P') then
        mark the frame as optional
    endif
    server = available
    while (queue ≠ {∅}) do
        if (frame is mandatory) then
            send the frame
            server = occupied
        else if (frame is hard optional) then
            if (all optional frames are rejected and
                its deadline is missed then
                reject the frame
                server = available
            else
                send the frame
                server = occupied
            endif
        else if (frame is optional) then
            if (its deadline is missed) then
                reject the frame
                server = available
            else
                send the frame
                server = occupied
            endif
        endif
    endwhile
end
```

In the following, are listed some advantages of the bandwidth fair sharing:

- To control the congestion of the network by fair sharing resources between all streams. A bad stream does not affect the service provided to the other streams, i.e. if a stream wants to consume more resources than available, only this service will be concerned.

- To guarantee an acceptable bandwidth and delay.

- To guarantee a link sharing between the different classes of service.

## D. k-frames: a method to (m,k)-frame guarantee

We define the concept of k-frames, which describes how a video stream composed of k frames is organized when deadlines are either met or missed. The k-frames of a stream with (m,k)-frame requirement is a succession of k elements from the alphabet Δ = {O,H,M} where:

- *O stands for a B frame;*
- *H stands for a P frame;*

- *M stands for an I frame.*

Using this specification, a video stream can express its (m,k)-frame constraint. In fact, the stream packets are labeled as optional, hard optional or mandatory according to their *k-frames*. To guarantee a minimum QoS of the stream, it is sufficient that all mandatory frames meet their deadlines, i.e. if some optional frames miss their deadline, then this leads only to the degradation of the (k,k)-frame QoS (hard guarantee), but does not affect the required (m,k)-frame QoS.

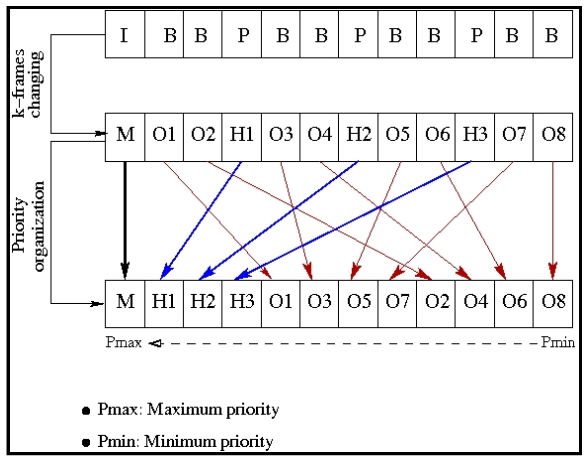

Figure 4. k-frames adaptation.

This timing constraint is very suitable for efficiently expressing the requirement of MPEG streams. In fact, an MPEG stream is organized in a cyclic GoP[4]. The frame types within a GoP have not the same importance. The loss of data in I-frames and/or in P-frames of an MPEG stream will be propagated and will cause errors in next frames until a new I-frame arrives, whereas the B-frame loss has no propagation effect. Hence, if a stream has a GoP structure *IBBPBBPBBPBB*, it could be considered as a stream with a (8,12)-frame requirement and it will be assigned a *k-frames* by *MOOHOOHOOHOO*, i.e. all B-frames are optional, whereas P-frames are hard optional and I-frames are mandatory. Therefore, scheduling processes must take more care of I frames since they are mandatory, then it must take more care of P-frames since they are hard optional. Let us illustrate this process with the figure 4.

Today's QoS architectures, especially FCA-DMS architecture, do not use the (m,k)-frame model for service guarantee. So, it would be useful to define a new QoS policy that integrates the (m,k)-frame guarantee offering a flexible way to express requirements of multimedia streams. Guaranteeing the in-time delivery of mandatory packets will provide a minimum acceptable QoS at the receiver end, and then, we would have a graceful degradation of QoS in overload conditions, i.e. where it is difficult to avoid packet losses.

When we apply (m,k)-frame classification and *k-frames* method to ensure the (m,k)-frame constraints on the GoP to the previous example, we obtain the result presented on Figure 5.

---

[4] Group of Pictures.

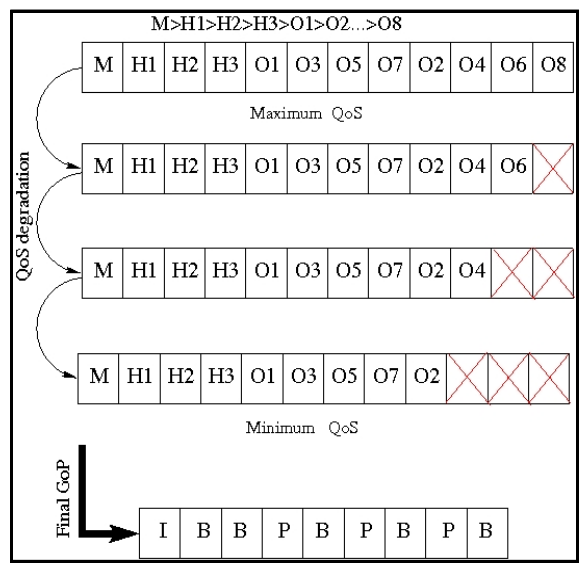

Figure 5. Final GoP after application of (9,12)-frame constraints.

## IV. REPLICATION STRATEGY

The role of the master server is to allocate a video server to a client with different manners. We are interested in the two next possibilities of attribution:

1) *The master server communicates the client address to a video server. Then, the stream becomes open to the client.*

2) *The master server communicates to the client the name of a video server and the client connects to it.*

### A. Specifications

After the video server has sent the video stream, the client receives the frames. In order to obtain a good visualization of video packets, each packet must be received by the client before a deadline. At the client side, packets must be received in real-time manner, i.e. they must respect time constraints.

The system must particularly minimize the value of the gap between two successive received packets. Thus, at frames level in each video packet, some temporal consistency must be respected. More precisely, the number of received frames per time unit must be proportional to the QoS required by the customer. Increasing and/or decreasing this number leads to a disruption of the received video and to the QoS degradation.

### B. Functionning and algorithm

A video server (VS) can only distribute videos stored on its disks. If a video is not accessible on several servers (only one VS contains this video), the probability that this VS becomes saturated increases. Therefore, it is necessary to define a new distribution strategy (cf. *Algorithm 2*) of video packets in order to have another video server which can be used to answer to the customer request. The saturated video server sends a request to its nearest video servers. Then, each video server is behaving according to one of the following three scenarios:

1) it possesses the video and it is able to treat the request (it is not saturated).
2) it possesses the video but it is unable to treat the request (it is saturated).
3) it doesn't possess the video, but it is probably able to treat the request because it is not saturated.

In the two first cases, the replication strategy is not established. In the last case, the case manager, that has to control replication, sends an order to the saturated VS to start the replication. Consequently, the case manager elects a VS among those that answered and that are not saturated. The choice of the VS is done in order to get the best possible QoS. The demand returns back again to the monitor, which then ends the replication process. Afterward, the monitor restarts.

```
vs:     video server;
ms:     master server;
cl:     client;
nvs:    neighbor of a video server;
snvs:   selected neighbor;
begin
   snvs={ø};
   vs is saturated and has the video;
   for all nvs_i do
      send-request(vs,nvs_i);
      if (nvs_i has not the video and not saturated) then
         snvs={ø};
         exit-for;
      else
         if (nvs_i has the video and not saturated) then
            put nvs_i in snvs;
         endif
      endif
   endfor
   if(snvs is not empty) then
      choose(nvs in snvs);
      video-replication(vs,nvs);
   endif
end
```

*Algorithm 2:* replication strategy algorithm

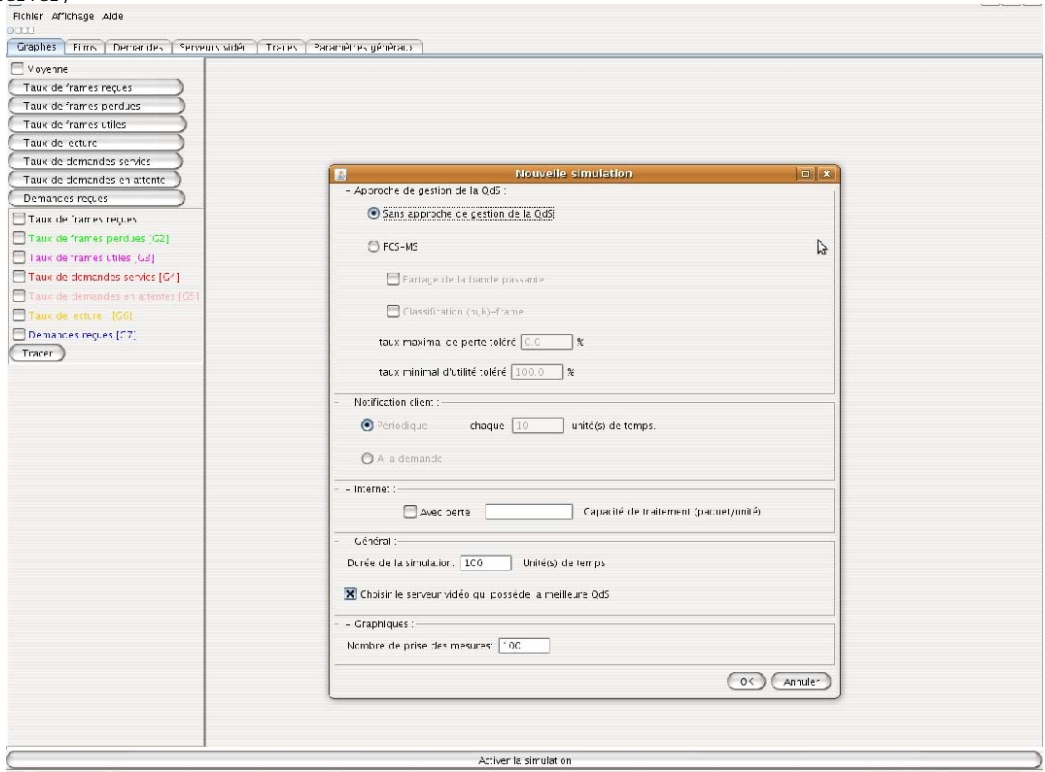

Figure 6. A simulator fors distributed multimedia systems.

## V. SIMULATIONS AND RESULTS

To assess the performances of (m,k)-frame method and replication strategy in comparison to previously proposed QoS approaches, we carried out simulations thanks to a simulator presented in Figure 6.

We studied the ratio behavior of the received client frames and the quality of service of the system. Given the main system parameters (described in Table I), we repeat the experiment 100 times in each simulation in order to obtain a sample of 100 values for the performances, i.e. to obtain significant results for QoS and rates. Each point showed in Figures 7 to 11 (rate of received frames, rate of useful frames, rate of lost frames, rate of waiting-frames and rate of served frames) represents the computed average of performance results deduced from each simulation sample.

### A. Presentation of simulations

To have significant periods of simulation and a huge number of measurements at every moment, is a difficult task for the system and the machine on which the simulation takes place. This can cause problems such as memory overflow. To deal with such problems, we fix the number of steps, then the system will calculate the time interval over which we average the measures (for instance, in a simulation of 10 000 time units, with 100 measurements, the simulator has to save the average value taken for a given parameter every *10 000/100* time units).

TABLE I. SIMULATION PARAMETERS.

| Notation | Definition | Value |
|---|---|---|
| **Characteristics of the system** | | |
| T-Sim | Time of simulation | 100 units |
| Nb-measure | Number of reported measures | [15,100] |
| **Characteristics of video stream** | | |
| Nb-video | Number of videos | [5,200] |
| Nb-GoP | Number of GoP per video | [20,100] |
| Nb-P | Number of P-frames per GoP | [3,9] |
| **Client demands** | | |
| $\lambda$ | Arrival rates of video streams | [0.1,2.0] |
| QoS | Quality of service (frames per unit of time) | [25,35] |
| Tm-service | waiting service | [1,20] |
| **Video server** | | |
| Nb-VS | Number of video servers | [5,100] |
| $\beta$ | Video server speed (frames per unit) | [100,200] |
| C | Video server capacity (number of videos) | [1,20] |

We assume that video streams arrive according to the *Poisson processes* of $\lambda$ parameter. The system load can vary depending on the number of video streams sent to the client. We assume that a transmission of a video packet requires an unit of measurement.

The workload of the system varrries according to both the load of video servers and the load of network. When $\lambda = 0.1$, the number of frames arriving to the network during one experiment is about 150 frames. When $\lambda = 2.0$, this number is about 1400 frames. And the network workload is related to the number of frames in the bandwidth.

The video server has an original role in distributed multimedia systems architecture. So, it is essential to allow the user of simulator to generate well-defined constraints. In fact, our simulator must offer the user the opportunity to perform this task in order to obtain a number of video servers having desirable characteristics.

In order to have a realistic aspect to our simulations, the Internet model is defined. This model allows the user:

- to integrate the concept of loss of frames in the network,
- to set the capacity of the Internet in a maximum number of frames processed per time unit.

### B. Simulations results

The objective of the simulations is to demonstrate how our method is able to adapt the QoS to the real conditions of a multimedia application, according to the current system load. Notably, the system must adjust the QoS when the number of clients carrying out requests varies, i.e dynamic arrival of clients. The network congestion can have different sources:

- *internal source:* when there is a large number of clients doing requests in the system. We can limit this number by using an admission controller located at the master server level.

- *external source:* when the network is used by other applications that can cause a congestion.

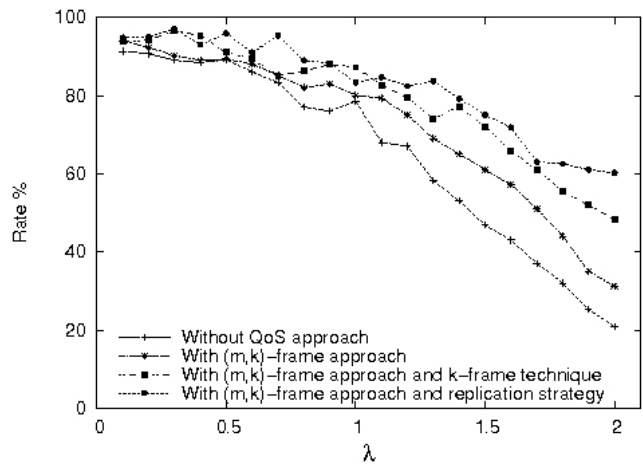

Figure 7. Rate of received frames.

In order to analyze the influence of the k-frames on the rate of received frames, we compare the results obtained when using the (m,k)-frame method and when varying the system workload. Figures 7 and 8 illustrate graphically this comparison.

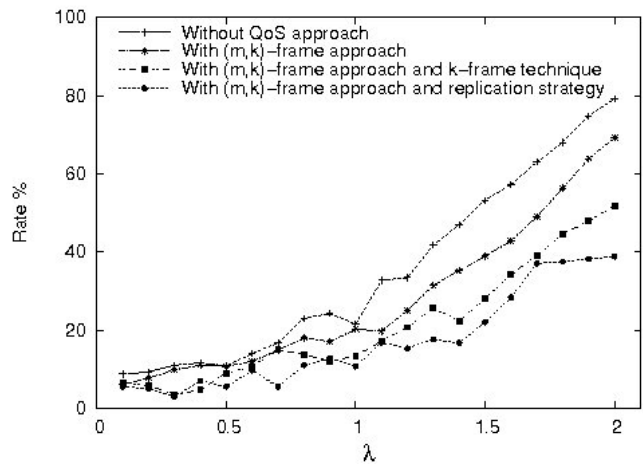

Figure 8. Rate of useful frames.

The best performances for transmitted frames are obtained when combining the (m,k)-frame method and *k-frames* technique (cf. Figure 7): the rate of received frame is the most important, i.e. 57%. We can see also in Figure 8 that for all variations of $\lambda > 0$, we obtain the best performances on the rate of useful frames and in rate of waiting-frames, i.e. for all system workload conditions. We can conclude that when increasing the load of transmitted frames, there is no great effect on the received frames, on useful frames and on waiting frames. This result may be explained by the higher priority assigned to mandatory frames (I), which ensures their processing before the other frames classes (P and B). When we look at the performances with *k-frames* technique (see Figure 7), we notice a progressive decreasing of the rate frames loss when the workload progressively becomes heavy.

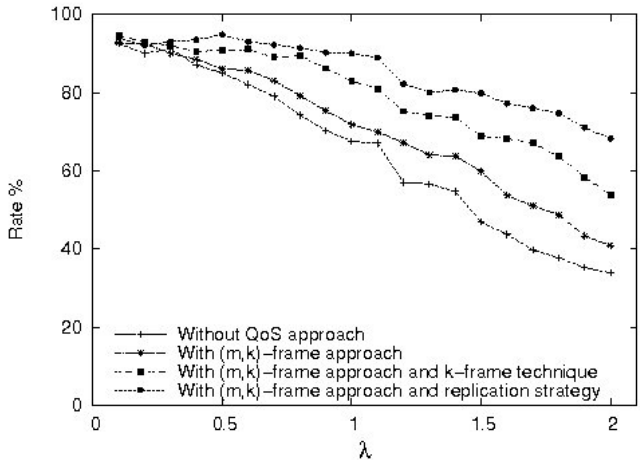

Figure 9. Rate of lost frames.

There is a difference when using only (m,k)-frame classification and when combining this method with the *k-frames* technique. In the latter case, P frames can be scheduled prior to B frames, which affects and decreases the rate of lost frames and degrades the served frames in the system (cf. Figures 9 and 11). This affects considerably the rate of received frames, especially when the system workload is heavy. In the following, we comment the *k-frames* performances on three intervals of λ, i.e. at different system workloads: λ ∈ [0.1, 0.7] (light workload), λ ∈ [0.8, 1.4] (average workload) and λ ∈ [1.5, 2.0] (high workload).

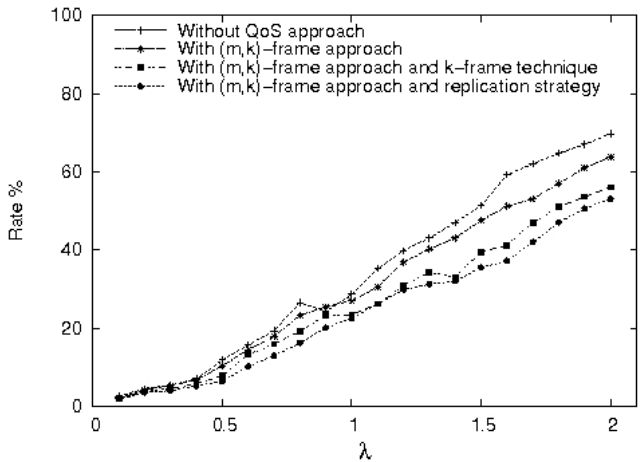

Figure 10. Rate of waiting frames.

When we consider the values of λ ∈ [0.1, 0.6] (cf. Figures 7 and 8), we notice that when the system is not overloaded, using both (m,k)-frame and *k-frames* methods give better performances on the rate of received and useful frames than using (m,k)-frame method only. Indeed, when using (m,k)-frame method, the lower priority frames (I) must wait for the execution of the higher priority frames (P and B). This has a negative effect when the system workload is light, which reduces the chances of lower priority frames to be transmitted.

When the system workload is average, i.e. value of λ is in the interval [0.7,1.4] (Figures 10 and 11), we can see that when combining (m,k)-frame and *k-frames* technique, we obtain better results than (m,k)-frame method only, according to the variations of the parameters of simulation.

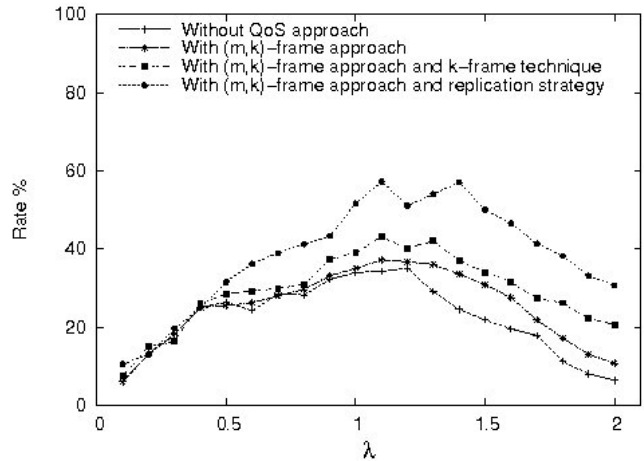

Figure 11. Rate of waiting frames.

When the system workload is heavy, i.e. λ ∈ [1.5, 2.0], the situation is reversed completely in favor of combining (m,k)-frame and *k-frames* technique, which provides better performances than the use of (m,k)-frame method only with all values assigned to the simulation parameters.

The results obtained by (m,k)-frame et *k-frames* techniques can be explained by the fact that k and m constraints are more guaranteed in this case than with only (m,k)-frame (Figure 10). We also deduce that when the workload increases, the improvement of the system performances is correlated with the addition of techniques used to control the QoS.

We noticed that after applying the replication strategy, we have obtained an important gain for the received frames and the useful frames. The replication strategy is an approach to resolve the congestion problem of video servers, whereas the (m,k)-frame method is used to resolve the problems of network congestion in overload situations.

In the following, we discuss and compare the system quality of service registered after applying the combination of (m,k)frame and *k-frames* techniques with those using the replication strategy of the video streams. Figures 7 to 11 illustrate graphically this comparison. When we look at the rate of received frames and useful frames (Figures 7 and 8), we deduce that all variants of workloads give the optimal performances on rate and QoS. We can conclude that adding a replication strategy has best effects on rates and gives high QoS on frames in all system workload conditions.

With (m,k)-frame and when adding the replication strategy, the waiting time of fresh frames is reduced thanks to the rate of received frames, which is important, i.e. 65%. This gives to the clients the maximum chances to meet their required QoS, decreasing then the system load. When the system workload is heavy, our approach reduces the lost frames, i.e. frames that are aborted and restarted by other video servers which are finally transmitted to the client.

## VI. CONCLUSION AND PERSPECTIVES

While current resource management systems provide mechanisms which provide reliability with respect to QoS, they seem to be not sufficient since there are many well established application scenarios where QoS adaptation is required, e.g., distributed multimedia systems. Our main contribution is related to the adaptation of a feedback architecture (FCS-DMS) to multimedia systems and the application a (m,k)-frame technique to them.

A possible extension of this work is the enhancement of the architecture that we have presented, in order to bring some fault tolerance because of the presence of only one master server.

We have also presented the importance of k-frames method to (m,k)-frame constraints guarantee and have given a high priority to I frames over P and B frames, in order to converge towards the QoS specified by the client. Simulations results allow us to validate the feasibility of our approach and allow to provide results demonstrating the real contribution of this new approach.

## AUTHORS PROFILE

**Bechir Alaya** is a PhD student at the Faculty of Sciences and Technics (University of Le Havre, France). His researches are done in two laboratories: MIRACL research group at ISIMS School (Tunisia) and LITIS laboratory (University of Le Havre). He received his Master degree in Computer Sciences from this university in 2007. His research interests include Quality of Service, Distributed Multimedia Systems and Feedback Control Architecture.

**Claude Duvallet** has obtained his PhD in October 2001. Since September 2003, he has been an associate professor at the University of Le Havre. His main topics of research are Real-Time Databases, Real-Time Systems and Multimedia Systems. He supervises many PhD students in these areas.

**Bruno Sadeg** is an associate Professor in the University of Le Havre (France). He is a head of a research team whose members work about "Intelligent Transport Systems". He is particularly interested by real-time mechanisms in sensor databases, embedded in vehicles and/or databases and located in some sites where vehicles get information, sometimes continuously.